\documentclass[copyright,creativecommons]{eptcs}
\providecommand{\event}{Wivace 2013} 

\usepackage{amsmath}

\usepackage{enumitem}
\mathchardef\mhyphen="2D

\newcommand{\mh}{\mhyphen}

\title{Application of a Semi-automatic Algorithm for Identification of Molecular Components in SBML Models}
\author{Andrea Maggiolo-Schettini
\institute{Dipartimento di Informatica\\
Universit\`{a} di Pisa, Italy}
\email{maggiolo@di.unipi.it}
\and Paolo Milazzo
\institute{Dipartimento di Informatica\\
Universit\`{a} di Pisa, Italy}
\email{milazzo@di.unipi.it}
\and Giovanni Pardini
\institute{Dipartimento di Informatica\\
Universit\`{a} di Pisa, Italy}
\email{pardinig@di.unipi.it}
}
\def\titlerunning{Identification of Molecular Components in SBML Models}
\def\authorrunning{A. Maggiolo-Schettini, P. Milazzo \& G. Pardini}
\begin{document}
\maketitle

\begin{abstract}
Reactions forming a pathway can be rewritten by making explicit the different molecular components involved in them.
A molecular component represents a biological entity (e.g. a protein) in all its states (free, bound, degraded, etc.).
In this paper we show the application of a component identification algorithm to a number of real-world models to experimentally validate the approach. Components identification allows subpathways to be computed to better understand the pathway functioning.
\end{abstract}

\section{Introduction}
%

A typical aspect of a biochemical pathway is that the process described involves mainly a few chains of reactions, in which the occurrence of a reaction produces some intermediate molecule which is then transformed subsequently by other reactions.
Just a few basic biological entities are often involved in a pathway, for example a simple protein undergoes a series of transformations, starting from its initial synthesized form, which can then be activated and also become part of different complexes. Therefore, intermediate molecular species can actually be seen as different states of the same initial biological entities. In accordance with this view, in the modelling of biochemical pathways we can consider a notion of \emph{molecular component} \cite{drabik2012,drabik2012conditions} that is the formal counterpart of the notion of biological entity. A molecular components will hence be associated with the set of species mentioned in the model and corresponding to different states of the same biological entity.

This paper discusses the application of a semi-automatic algorithm for the identification of molecular components in pathways which is presented in~\cite{PardiniMMS13-components}. 
The algorithm infers components from the interactions of molecular species. In order to make components involved in each reaction explicit, the algorithm replaces each compound species (which is usually identified by a distinct name) with a fresh name for each one of its components, and transforms reactions into a normal form in which each complex is represented by as many species as are the biological entities involved in it.
The algorithm is actually semi-automatic, since there are cases in which the components involved in a reaction cannot be univocally determined from the context; these cases must be solved with the help of the user.

In this paper, we present the results of the application of such an algorithm to a number of real-world pathways from the BioModels repository~\cite{BioModels2010}, an online database of machine-readable models of biological processes formalized in the well-known Systems Biology Markup Language (SBML).
In particular, we have analyzed all of the models in the curated section, which contains only models from the literature, for a total of 436 models. In this paper, some of the models are also discussed individually in order to experimentally assess the validity of our approach when applied to real-world models.

The identification of components by the transformation of the pathway reactions into normal form allows different kinds of analysis to be performed.
For example, syntactic transformations, such as the projection of the pathway over a subset of components, would allow the user to obtain insights on the functioning of the pathway. In particular, by focusing on the reactions in which they are involved, the mechanisms underlying the pathway dynamics could be more easily identified.
Moreover, from a theoretical point of view, the normalization would allow the automatic translation of a pathway into a set of automata or terms of process calculi \cite{phillips2007efficient,barbuti2008calculus,barbuti2006calculus,ciocchetta2009bio,danos2004formal}, thus enabling the use of any tool available for their analysis.

\section{Algorithm for Component Identification}
In this section we recall the algorithm for component identification presented in~\cite{PardiniMMS13-components}.
The algorithm is based on the idea that a species is a ``state'' of a more abstract biological entity, and a reaction is a synchronized state change of a set of such entities. The algorithm assumes that biological entities involved in a reaction cannot appear from nothing and/or disappear (i.e. degradations should be modelled by using a species representing the degraded state of the biological entity). As a consequence, every entity involved in a reaction should be mentioned both among reactants and among products, and hence in a reaction there should be as many reactants as products. 

Let a {\em component} be a set of species representing all possibile states of a given biological entity. The component identification algorithm transforms a given pathway into a \emph{normal-form} pathway, namely such that, for all its reactions, there are as many reactants as products and there is a one-to-one positional correspondence between reactants and products which are part of the same component. The fundamental operation of the algorithm is to split a species into new subspecies, denoted by newly-introduced symbols. A split of a species is performed any time the algorithm can infer that, in a reaction, the occurrence of such a species must actually denote a complex composed of multiple bound molecules. For example, this is the case of a reaction $A,B \to C$, where $C$ can be split into subspecies $C_A, C_B$, obtaining the normal-form reaction $A,B \to C_A,C_B$,
with components $\{A,C_A\}$ and $\{B,C_B\}$.
In general, this process may not be completely automatic since some ambiguities may arise. For example, a reaction such as $A,B \to C,D,E$ is ambiguous since it is not clear which of the two reactants has to be split into two in order to obtain as many reactants as products.

Another important aspect is that, in principle, the number of components of a pathway is not univocally determined.
For example, it is possible to split the reactants/products of a reaction $A \to B$ into $A_1,\ldots,A_n \to B_1,\ldots,B_n$, thus identifying any number of components, one for each pair $A_i,B_i$.
However, the idea is to identify only those components which can be inferred from the context, namely from the reactions in which a component is involved.
For this reason, the algorithm performs a split only if it is needed to match the components of reactants and products of a reaction. 



\smallskip \noindent {\bf Structure of the algorithm}
Let us denote (in an abstract manner) by $comp(S)$ the component in which a species $S$ occurs at a given step.
Initially, the algorithm assumes that each species occurring in the input pathway is part of a different component.
It then iteratively performs two alternating phases, until a normal-form pathway is obtained.
In the first phase, it tries to transform the set of reactions into reactions having the same number of reactants and products. This entails inspecting each reaction in turn, in order to both (i) split species into subspecies as needed, and (ii) refine the information about components by collapsing different components into single ones, according to what can be inferred from reactions.

In case of ambiguities, the algorithm may fail to generate a normal-form pathway since either the number of reactants and products differ, or the correspondence between reactants and products is not completely specified for all the reactions. In such a case, the algorithm performs a second phase, which demands user intervention to resolve a single ambiguous reaction. Then the two phases are repeated to propagate the new information derivable from the resolved reaction to the others, as it may be useful to resolve other ambiguous reactions, and possibly all of them.

As regards the first phase, the algorithm examines each reaction singularly.
Let us consider a reaction having reactants $R = \{A_1, \ldots, A_\gamma\}$ and products $P = \{B_1,\ldots,B_\delta\}$.
Moreover, assume that the components of each of the first $k < \gamma,\delta$ pairs of reactant/product match, namely $\forall i \in \{1,\ldots,k\}.\ comp(A_i) = comp(B_i)$.
Let $n=\gamma - k$ and $m=\delta - k$ denote the number of remaining unmatched reactants/products of the reaction, respectively.
The algorithm then distinguishes five kinds of reactions, as follows:
\begin{itemize}[noitemsep,nolistsep]
 \item Case $n=0$, $m=0$: the reaction is resolved, with complete correspondence between each reactant and product.
 This case is not necessarily definitive, since the algorithm, due to some split, may subsequently modify the reaction by replacing a species occurring in it with new symbols.
 \item Case $n=1$, $m=1$: since there is just one species remaining in each side, this means that both of them are part of the same component. Thus, the components of the reactant and product are joint into a single component by taking their union.
 \item Case ($n=0 \land m>0$) or ($n>0 \land m=0$): if either of reactants or products are empty, there is at least a component which appears in either side of the reaction which does not occur in the other side. In other words, such a component either appears or disappears in between the reaction, which is not allowed by our approach. Thus this case is regarded as an \emph{error}, which cannot be resolved.
 \item Case $n=1 \land m>1$ (or $n>1 \land m=1$): in this case there is a single reactant which is actually formed from $m$ subspecies. 
 Thus, $m$ new symbols $A_\gamma\mh B_k, \ldots, A_\gamma\mh B_{\delta}$ are introduced, by combining the name of $A_\gamma$ with the $B_i$'s. Then $A_\gamma$ is replaced by the new symbols in all the reactions, and the components of each pair of species $(A_\gamma\mh B_k, B_k), \ldots, (A_\gamma\mh B_\delta, B_\delta)$ are matched, i.e. $\forall k \leq j \leq \delta.\ comp(A_\gamma\mh B_j) = comp(B_j)$. (The converse case, with one product and $n$ reactants is analogous.)
 \item Case $n>1$, $m>1$: if there are multiple unmatched reactants and products, the reaction is currently \emph{ambiguous}. This means that the current information regarding components does not allow the algorithm to decide which species must be split, thus this reaction is skipped.
\end{itemize}
The first phase continues until there are only resolved, erroneous and ambiguous reactions.
In case of ambiguous reactions remaining, the algorithm asks the user to resolve one reaction by specifying the species to split and rewriting it in normal-form. The procedure is repeated from the first phase in order to propagate the updated information about components to the other reactions.

\newcommand{\Ga}{G_{\alpha}}
\newcommand{\Gbc}{G_{\beta\gamma}}
As an example, let us consider pathway
\begin{small}
$P=\{\mathbf{r_1}: Lig, rcpt \to C_1;\ 
\mathbf{r_2}: GDP, \Ga \to C_2;\ 
\mathbf{r_3}: GTP, \Ga \to C_3;\ 
\mathbf{r_4}: C_3 \to C_2;\ 
\mathbf{r_5}: C_2, \Gbc \to C_4;\ 
\mathbf{r_6}: C_4, C_1 \to C_5\}$%
\end{small},
which models a small fragment of the well-known G protein signalling pathway,
in which a ligand {\small$Lig$} (representing the signal) binds the receptor {\small$rcpt$} on cellular membrane, triggering the internal process.
The algorithm identifies $5$ components (representative species of which are
$Lig$, $rcpt$, $\Ga$, $\Gbc$ and $GDP$) and produces the follwing normal-form pathway (where species names are abbreviated):
\begin{small}
\begin{align*}
P' = & \{\mathbf{r_1'}: Lig, rcpt \to C_1\mh Lig, C_1\mh rcpt;\ 
\quad \mathbf{r_2'}: GDP, \Ga \to C_2\mh GDP, C_2\mh \Ga;\ 
\quad \mathbf{r_3'}: GTP, \Ga \to C_3\mh GTP, C_3\mh \Ga;\\ 
& \phantom{\{} \mathbf{r_4'}: C_3\mh GTP,\allowbreak C_3\mh \Ga \to C_2\mh GDP, C_2\mh \Ga;\ 
\quad \mathbf{r_5'}:
      C_2\mh GDP, C_2\mh \Ga, \Gbc \to
      C_4\mh GDP, C_4\mh \Ga, C_4\mh \Gbc;\\ 
& \phantom{\{} \mathbf{r_6'}:
      C_4\mh GDP, C_4\mh \Ga,\allowbreak C_4\mh \Gbc, C_1\mh Lig, \allowbreak C_1\mh rcpt \to C_5\mh GDP, C_5\mh \Ga, C_5\mh \Gbc, C_5\mh Lig, C_5\mh rcpt\}.
\end{align*}
\end{small}
The sets of species forming each component are as follows:%
\begin{small}
$\{Lig, C_1\mh Lig, C_5\mh Lig\},
\{rcpt, C_1\mh rcpt, C_5\mh rcpt\},\linebreak 
\{\Ga, C_2\mh \Ga, C_4\mh \Ga, C_5\mh \Ga, C_5\mh \Ga, C_3\mh \Ga\},
\{\Gbc, C_4\mh \Gbc, C_5\mh \Gbc\},
\{GDP, C_2\mh GDP, C_4\mh GDP, C_5\mh GDP, GTP,\linebreak  C_3\mh GTP\}$.
\end{small}
Now, we can compute subpathways describing the activity of a subset of the molecular components. For instance, by assuming that we are not interested in the role of $Lig$, $rcpt$ and $GDP$, we can compute the subpathway dealing only with components $\Ga$ and $\Gbc$ as follows:
\begin{small}
$P'' = \{\mathbf{r_2''}: \Ga \to C_2\mh \Ga;\ 
\mathbf{r_3''}: \Ga \to C_3\mh \Ga; \mathbf{r_4''}: C_3\mh \Ga \to C_2\mh \Ga;\ 
\mathbf{r_5''}: C_2\mh \Ga, \Gbc \to C_4 \Ga, C_4\mh \Gbc; \mathbf{r_6''}: C_4\mh \Ga, C_4\mh \Gbc \to C_5\mh C_2\mh \Ga, C_5\mh \Gbc\}.$
\end{small}
Moreover, the subpathways describing each component individually can be trivially translated into a finite state automaton or into a process algebra term to enable the application of formal analysis tools.

\section{Applications}

In order to test our component identification algorithm on a relevant number of real pathways we downloaded all of the pathway descriptions available in the BioModels database \cite{BioModels2010} under the category ``curated models''. So, our testbed consisted of 436 different SBML models of pathways.

SBML \cite{SBML03} is a well-established XML-based language for pathway description. A SBML pathway model includes (among others) the following elements: 
\begin{itemize}[noitemsep,nolistsep]
 \item {\bf Species:} Proteins, genes, ions and other molecules that can participate in reactions.
 \item {\bf Compartments:} Well-stirred containers in which species can be located. A SBML model may contain multiple compartments (at least one) and each species must be located in one of them.
 \item {\bf Reactions:} Statements describing transformation, transport or binding processes that can change the amount of one or more species. A reaction consists of reactants, products and modifiers. Moreover, the kinetic law of a reaction can be expressed by using arbitrary mathematical functions.
 \item {\bf Rules:} Mathematical expressions describing how some variable values (e.g. species amounts) can be calculated from other variables, or used to define the rate of change of variables. Rules in a model can be used either together with (or in place of) reactions to determine the model dynamics.
 \item {\bf Events:} A statement describing an instantaneous, discontinuous change in a set of variables when a triggering condition is satisfied.
\end{itemize}
We developed a simple translator of SBML models into CSV files accepted by our implementation of the algorithm. The translator considers only the reactions of a SBML model and transforms them into the format expected by the algorithm. SBML models in which the dynamics is governed only by rules, and not by reactions, are translated into empty CSV files, and hence are unusable by our tool. Moreover, we excluded from our test models containing reactions with fractional stoichiometry, that are meaningful only when the dynamics is described by means of ODEs. In the considered testbed, unusable models turn out to be 59 out of 436 and excluded models are 23 out of 436, hence we have 354 usable CSV files.

The 354 usable models consisted on average of $22.07$ species and $31.05$ reactions, with 108 models dealing with more than 20 species, 31 dealing with more than 50 species, and 6 dealing with more than 100 species. We executed our algorithm batch on all of the models. The execution (on a standard laptop) takes a few seconds for each model. The execution on a model can terminate either successfully with an automatically generated normal-form pathway (OK), or with an error message if an erroroneous reaction is encountered, or with a human intervention request if an ambiguous reaction is encountered. In the latter two cases, the execution is interrupted and the number of erroneous and ambiguous reactions, respectively, is printed. The results we obtained on the 354 usable models are shown in the table below, on
line ``Batch execution''. Most of the models (244 out of 354) encountered some errors. By inspecting some of them we 
discovered that in most cases the error was due to reactions describing either synthesis or degradation of some species, in which either products or reactants were absent, respectively.

\small
\begin{center}
\begin{tabular}{|r|c|c|c|}
\hline
 & Ok & Erroneous & Ambiguous\\
\hline 
Batch execution & 96 & 244 & 14\\
Preprocessing \& Batch execution & 241 & 93 & 20\\
Preprocessing \& Batch execution with dynamic correction & 318 & 0 & 36\\
\hline
\end{tabular}
\end{center}
\normalsize

We decided to solve the problem of synthesis/degradation reactions by preprocessing the SBML models. The preprocessing was performed on the CSV translation and consisted in inserting a dummy species in each empty set of reactants or products encountered. Note that every time an empty set is encountered by the preprocessor, a new fresh species is generated and used as dummy. Hence, each dummy species will appear only once in the pathway. Since dummy species added during preprocessing do not interact with other species, error situations are solved without affecting component identification.

The results of executions of the algorithm after preprocessing are in the table, on
line ``Preprocessing \& Batch execution''. Most of the previously erroneous situations can now be handled by the algorithm, and in the vast majority of the cases they did not encountered ambiguous reactions (151 errors are solved and only 6 of them turn out to need human intervention).

By inspecting some of the models still giving an erroneous result we discovered that in some cases the error was caused by a reaction describing the degradation of a part of a complex. For example, let us consider a pathway consisting of reactions $A,B \to C$ and $C \to A$. The first reaction is the formation of a complex composed by $A$ and $B$, and the second is the degradation of $B$ when it bound to $A$. In this case the algorithm transforms the first reaction into $A,B \to C_A,C_B$. Consequently, in the second reaction it replaces $C$ by $C_A,C_B$ obtaining $C_A,C_B \to A$. Now, in the second reaction the algorithm associates $C_A$ with $A$ (as this follows from the first reaction) and it has nothing to associate with $C_B$ (error situation).

To solve this second kind of errors we choose to add fresh dummy objects also at runtime when errors are encountered. In the example, the second reaction could be corrected by adding a fresh dummy object $D_B$ as follows: $C_A,C_B \to A,D_B$.
This is not always correct since the dynamic insertion of dummy species may lead to different components being identified, depending on the order of processing of reactions.
The results of executions of the algorithm after preprocessing and with insertion of dummy objects at runtime are in the table, on
line ``Preprocessing \& Batch execution with dynamic correction''. All of the error situations are now solved and only 36 of the 354 models turn out to need human intervention.

The component identification algorithm completed automatically its execution in the 89.83\% of the cases. In the remaining 36 cases we had an average of 4.67 ambiguous reactions, that means that at most 4.67 questions are asked to the user in an average execution of a model with ambiguous reaction (with a maximum of 16 and only two models over 10). The average computed over all of the 354 cases is 0.53.

The algorithm, although semi-automatic, turns out to require very limited human intervention in practice. Let us now assess the quality of the computed results by comparing the molecular components identified by the algorithm with the biological entities the modelled pathways deal with according to the referenced literature. We consider eight ``randomly'' chosen SBML models (numbers 50, 100, 150, 200, 250, 300, 350 and 400 in the BioModels database), three randomly chosen big models (numbers 88, 235 and 293) and three randomly chosen models with ambiguities (numbers 82, 143 and 165). In the following paragraphs a brief summary of the analysis of each of these SBML models is given.

\smallskip

\noindent {\bf Analysis of model 50 (BIOMD0000000050.xml).} This SBML model describes a kinetic model of N-(1-deoxy-D-fructos-1-yl)-glycine (DFG) thermal decomposition \cite{martins2003kinetic}. The SBML model includes 14 species and 16 reactions. During its execution the algorithm encountered one erroneous reaction, that was solved by dynamically inserting a dummy object in it. In the model DFG can be degraded into several different ways obtaining a number of different substances. The model includes some species representing unidentified intermediate components ($E_1$ and $E_2$) and unidentified carbohydrate fragments ($C_n$). Most of the species involved in the pathway can be transformed into these unidentified molecules, thus causing all of these species to be included in the same molecular component by the algorithm. As a consequence, the algorithm identifies only 3 components: two including all and only the species related with Glycine and methyulglyoxal, respectively, and one including all of the 
other species. The  quality of the result in this case are hence only partially satisfactory, and the cause of unsatisfaction is that the model includes ambiguities (unidentified species).

\noindent {\bf Analysis of model 82 (BIOMD0000000082.xml).} This SBML model describes the formation of an inhibitor of the Adenylate Cyclase enzyme \cite{thomsen1988inhibition}. The SBML model includes 10 species and 6 reactions. During its execution the algorithm encountered no erroneous reactions, but two ambiguous reactions. The two ambiguous reactions are the following: $DR, G\_GDP \to DRG\_GDP$ and $DRG\_GDP \to GDP, DRG$. These reactions describe the passage (in two steps) of a G-protein from $GDP$ to $DR$. This is a typical case of ambiguity (see \cite{PardiniMMS13-components}) that can be solved by making it explicit the involvement of the ``hidden'' component representing the $G$ protein, namely by replacing the ambiguous reactions with the following ones: $DR, G_{G\_GDP}, GDP_{G\_GDP} \to DRG\_GDP$ and $DRG\_GDP \to GDP, G_{DRG}, DR_{DRG}$. So, in this case the algorithm needs human intervention, after which 4 components are correctly identified.

\noindent {\bf Analysis of model 88 (BIOMD0000000088.xml).} This model describes the Rho-kinase pathway in order to study thrombin-dependent in vivo transient responses of Rho activation and Ca$^{2+}$ increase \cite{maeda2006ca2+}. It includes 104 species and 110 reactions, and it is one of the biggest models we have considered. During its execution the algorithm encountered no erroneous and ambiguous reactions. The algorithm identified 28 components. Species in the SBML file are represented by numbers, hence checking the correctness of the identified components was not trivial. We checked a few randomly chosen components and each of them turned out to be composed of species representing different states of the same molecule. Although we do not have a complete proof of the correctness of the result, we consider this case satisfactory.

\noindent {\bf Analysis of model 100 (BIOMD0000000100.xml).} This SBML model is used to study the effects of cytosolic calcium oscillations on activation of glycogen phosphorylase \cite{rozi2003theoretical}. The SBML model includes 5 species and 10 reactions. During its execution the algorithm encountered five errors, that were solved by preprocessing. The species involved in the model are: external calcium (EC), cytosolic calcium (Z), intravescicular calcium (Y), inositol 1,4,5-trisphosphate IP$_3$ (A), and glycogen phosphorlylase (GP). The reactions described in the model are: calcium influx, transportation of calcium between compartments, and IP$_3$ and glycogen phosphorylase syntheses and degradations. The algorithm identifies 3 components: one including all of the species representing calcium (EC, Z and Y) and the other two including IP$_3$ and glycogen, respectively. The quality of the result in this case is hence satisfactory.

\noindent {\bf Analysis of model 143 (BIOMD0000000143.xml).} This SBML model describes the metabolism of activated neutrophils in which oscillatory behaviours have been observed \cite{olsen2003model}. The model includes 20 species and 20 reactions. During its execution the algorithm encountered three errors, solved by preprocessing. The algorithm encountered also 9 ambiguous reactions. After inspecting the model we discovered that thee reactions in particular were problematic, the solution of which solved also the ambiguities in the other 6 reactions. The three problematic reactions describe the {\em Peroxidase Cycle} in which an enzyme is activated by a $H_2O_2$ molecule and then transforms in two steps two Melatonin molecules into Melatonin-free-radical. During this cycle two molecules of water are released, and also some hydrogen ions are involved. Water and hydrogen are not mentioned in the model, and this creates ambiguities in the reactions. Such ambiguities can be  solved by a 
human intervention aimed at clarifying the three problematic reactions. After this, the algorithm completes by correctly identifiying 5 components.

\noindent {\bf Analysis of model 150 (BIOMD0000000150.xml).} This is a very simple SBML model describing formation and activation of Cdk/Cyclin Complexes \cite{morris2002kinetic}. The model includes 4 species and 4 reactions. The algorithm encountered no erroneous reactions. The modelled reactions are trivial (complex formation from two species, activation of the complex and inverse reactions). The algorithm correctly identifies 2 components, one for each molecule involved in the complex. The result in this case is hence satisfactory.

\noindent {\bf Analysis of model 165 (BIOMD0000000165.xml).} This model deals with intracellular signalling through cAMP and its cAMP-dependent protein kinase (PKA) \cite{saucerman2006systems}. The SBML model includes 37 species and 30 reactions. During its execution the algorithm encountered three erroneous reactions, two of which solved by preprocessing and one by dynamically inserting a dummy object. The model included one ambiguous reaction similar to those encountered in the case of model 82. The reaction can be disambiguated by means of a single human intervention after which 16 components are correctly identified.

\noindent {\bf Analysis of model 200 (BIOMD0000000200.xml).} This SBML model describes binding reactions leading to a transmembrane receptor-linked multiprotein complex involved in bacterial chemotaxis \cite{bray1995computer}. The SBML model includes 21 species and 34 reactions. During its execution the algorithm encountered no erroneous reactions. The biochemical entities involved in the pathway are the dimeric aspartate-binding receptor Tar (TT) and four cytoplasmic proteins CheW (W), CheA (AA), CheY (Y), CheB (B) and CheZ (Z). TT, W and AA are involved in a number of reactions leading to the formation of a complex TTWWAA that activates proteins Y, B, Z in different ways . The algorithm identifies exactly 6 components out of the 21 species, and such components correspond exactly to the 6 biochemical entities involved in the pathway. The quality of the result in this case is hence satisfactory.

\noindent {\bf Analysis of model 235 (BIOMD0000000235.xml).} This SBML model describes the sea urchin endomesoderm network \cite{kuhn2009monte}, a very big gene regulation network. The model includes 618 species and 778 reactions. During its execution the algorithm encountered only 3 erroneous reactions, solved by preprocessing. The algorithm identified only 47 components. By inspecting them we discovered that actually one of the components included the vast majority of the species. By inspecting reactions we discovered that actually they consisted mostly of syntheses and degradations in which species ``none'' was used in all the case of empty reactants or products. The algorithm associated all of the species involved in a synthesis or degradation reaction in the same component. This was due to the fact that all of such species are in relation with the same species ``none''. We modified the model by replacing the unique ``none'' by fresh dummy species (as in the case of preprocessing). After this  change the algorithm identified 406 components. We checked some of them (randomly chosen) and they turned out to be correct representations of biochemical entities involved in the network. We consider hence the quality of the result satisfactory, although this model needed to be slightly modified in order to let the algorithm work as expected.

\noindent {\bf Analysis of model 250 (BIOMD0000000250.xml).} This SBML model is used to study how epidermial growth factor (EGF) and heregulin (HRG) generate distinct responses of the transcription factor c-fos \cite{nakakuki2010ligand}. The SBML model includes 49 species and 78 reactions. During its execution the algorithm encountered 19 erroneous reactions solved by preprocessing. The algorithm identifies 18 components that, after an analysis of the description of the pathway in the paper, seem to correspond to the biochemical entities involved in the pathway. The quality of the result in this case is hence satisfactory.

\noindent {\bf Analysis of model 293 (BIOMD0000000293.xml).} This SBML model describes an ubiquitin-proteasome system \cite{proctor2010modelling}. The SBML model includes 136 species and 316 reactions. During its execution the algorithm encountered 114 erroneous reactions, all solved by dynamic insertion of dummy objects. The algorithm identified only 12 components. By inspecting the model we discoverd that is suffered from the same problem of model 235, namely the same dummy species ``source'' and ``AggP\_Proteasome'' were used in many syntheses and inhibition reactions, causing the most of the species to be assigned to the same component by the algorithm. We modified this model as we did for model 235. The number of erroneous reactions encounterd decreased to 77, and the number of components identified by the algorithm decreased to 16. Components seems to be rather correct, since they seems to represent a reasonable partition of the species set. However, the complexity of the model and the high number 
of erroneous reactions solved only at runtime does not allow us to be sure about the correctness of the result. We leave the assessment of the quality of this result as a future work.

\noindent {\bf Analysis of model 300 (BIOMD0000000300.xml).} This SBML model is used to study how the activity of the heterodimeric transcription factor hypoxia inducible factor (HIF) if affected by the interaction of factors inhibiting HIF (FIH) with ankyrin-repeat domain (ARD) proteins \cite{schmierer2010hypoxia}. The SBML model includes 9 species and 10 reactions. Reactions are all degradations and syntheses of species. All of the 10 reactions become erroneous reactions that are solved by preprocessing. The algorithm identifies 9 components. However, by inspecting the models it emerged that most of the dynamics is described by means of SBML rules, hence the models turns out to be unusable.

\noindent {\bf Analysis of model 350 (BIOMD0000000350.xml).} This SBML model is used to study the circadian network of higher plants. In particular, it is a model of the clock of the picoeukaryotic alga Ostreococcus tauri as a feedback loop between the genes TOC1 and CCA1 \cite{troein2011multiple}. The SBML model includes 14 species and 30 reactions. Actually, most reactions are syntheses and degradations. Indeed, 27 reactions out of 30 become erroneous reactions that are solved by preprocessing. Only three reactions describe transformation of species, and consequently the algorithm correctly identifies 11 components. The quality of the result in this case is hence satisfactory.

\noindent {\bf Analysis of model 400 (BIOMD0000000400.xml).} This SBML model belongs to the set of unusable models since it does not include any reaction.

\section{Discussion}

The component identification algorithm we proposed in \cite{PardiniMMS13-components} turned out to work pretty well. In the majority of the cases the execution of the algorithm has been completely automatic, and the computed molecular components correctly represented the biological entities involved in the pathway. In the cases in which human intervention was necessary, it usally consisted in answering very few questions on how resolve ambiguous reactions. After the detailed analysis of some of the models (numbers 235 and 293) a recurrent ``error'' in the modelling of pathways (from the viewpoint of the algorithm) is to use the same special species to represent the pre-synthesis or the degraded form of many different biological entities. This causes different molecular components to be erroneously merged into one. Moreover, it also emerged that ambiguities often are of the same kind, namely they are include some ``hidden'' species that is always bound to some other species. In these cases the algorithm 
cannot 
identify the component corresponding to such species and the reactions turn out to be ambiguous.

Most of the erroneous and ambiguous situations could be prevented by a more accurate construction of models. However, it could be an interesting further development of our work the definition of some rule or heuristics able to solve these situations when encountered. Otherwise, the prevention of erroneous and ambiguous situations could be approached by developing of a model repair preprocessing routine based for instance on static checking of conservation laws or P-invariants \cite{clark2012formal}.

\nocite{*}
\bibliographystyle{eptcs}
\bibliography{compdemobib}
\end{document}